# Perspective on Code Submission and Automated Evaluation Platforms for University Teaching

Florian Auer[a], Johann Frei[a], Dominik Müller[a], Frank Kramer[a]

[a]*IT-Infrastructure for Translational Medical Research, University of Augsburg, Augsburg, Germany*

## Abstract

We present a perspective on platforms for code submission and automated evaluation in the context of university teaching. Due to the COVID-19 pandemic, such platforms have become an essential asset for remote courses and a reasonable standard for structured code submission concerning increasing numbers of students in computer sciences. Utilizing automated code evaluation techniques exhibits notable positive impacts for both students and teachers in terms of quality and scalability. We identified relevant technical and non-technical requirements for such platforms in terms of practical applicability and secure code submission environments. Furthermore, a survey among students was conducted to obtain empirical data on general perception. We conclude that submission and automated evaluation involves continuous maintenance yet lowers the required workload for teachers and provides better evaluation transparency for students.

*Keywords:*
Teaching, automated code evaluation, education, student assignment management

## Introduction

Acquiring programming proficiency still remains a matter of continuous practice, and professional guidance remains a key aspect for success. In-person sessions allow teachers to react to the individual issues of students and thus enable better progressions. Therefore guided practical courses on basic and advanced topics play a major role in teaching at universities [1,2].

A crucial bottleneck in practical courses is the submission and evaluation of programming code of the course participants. Submissions of solutions for tasks and sub-tasks are usually the basis for deriving grades, and therefore need to be documented. An often seen method is either sending the code by email or copying files to a USB flash drive, followed by a manual review by a tutor or teacher. For obvious reasons, this method is not only time-consuming and error-prone but also tedious, and can easily be undertaken by a platform with included automated evaluation [3,4].

Because of the current COVID-19 pandemic, the face of teaching changed rapidly from in-person meetings to virtual sessions [5]. Online tools now have to fill the gap left behind by not being able to give direct feedback on problems and errors a student might encounter while solving programming tasks. Providing the utmost verbose feedback is crucial for supporting the students in their progression [6].

In this work, we describe a perspective on how a submission and automated evaluation platform can help to facilitate university teaching, especially in practical courses.

## Methods

For the submission and automated evaluation (SAE) of programming code, we developed our platform MISITcms. It is based on an existing contest management system called CMS [7], which development is discontinued.

On our platform, it is possible to define several tasks for which the students can upload their solutions (Figure 1). To determine the correctness of the submissions, certain checks can be implemented, ranging from a simple comparison of the output of the run scripts to more comprehensive tests performed on the in- and output data.

A simple case would be to run the submitted scripts with different parameters and compare the output with a predefined text file. While this is already sufficient for most cases, it has its limitations if the result of the program is dependent on changing input data or has a random component to it. In those cases, it is necessary to specify a more sophisticated evaluation of the results.

In both cases, it is possible to report to the student under which circumstances the submission fails and which errors occur while running their code. This feature not only is valuable to the students but also reduces the time and effort teachers have to invest in testing.

Running external code on a local machine always includes security issues. To reduce and even avoid potential risks, the submitted source code is evaluated in a secure environment. On our platform, this is realized using Isolate containers [8,9]. Although we taught Python [10] in our programming courses, the secure environment is not limited to this programming language. Multiple languages are supported due to dynamic programming language interfaces.

Executing code in a sandbox comes with some downsides: Required input data must be mounted within the containers to be accessible for the scripts. Furthermore, the results, i. e. the output of run scripts and occurring errors have to be stored not only for documentation purposes but also to provide the previously mentioned feedback for the students. To accomplish this, it is possible to define specific directories to be mounted within the sandbox on a task-specific level.

*Figure 1 – Interface for students showing the different exercises, upload form for solutions and results for previously evaluated submissions.*

For security reasons, the isolation of the sandbox also includes that internet access is locked from within the containers. However, in some cases, access to online resources is required, for example, if some assignments include the usage of certain APIs to retrieve online data. For those cases, it possible to grant networking permissions on a task-specific level to allow internet connections inside the Isolate container.

Third-party modules, libraries, or frameworks are often required for solving complex tasks. Thus, our platform supports specifying predefined lists of libraries and modules for certain tasks. Those are automatically loaded and integrated into the container environment when the tasks are evaluated.

Providing a platform for code evaluation is also limited by the available server-side resources. Each submission has to be evaluated, and with an increasing number of students also the demands on computational capabilities are growing. Basic approaches for executing submissions in parallel tend to fail and lead to more sophisticated queuing strategies. To surpass those limitations, it is possible to define several workers, that are responsible for evaluating one task (Figure 2). Submissions are queued and distributed over those workers to balance the load of the system. In the admin panel, the teachers have an overview of running and scheduled tasks and even can activate and disable available workers.

Proper practical courses usually take place over several weeks or months, with a significant amount of lecture slides, exercises, example code and code fragments, and different kinds of data provided. Unfortunately, it is not unusual that those a scattered across several platforms. Our platform, therefore, acts as a hub to consolidate the different documents for the students. To not overwhelm the students with the huge number of files at once, it is also possible to unlock those files needed for the current and previous days.

The setup of such kind of a platform tends to be not quite simple in terms of software dependencies need to be installed and databases have to be prepared and structured. To simplify this process for others, we provided a dockerized [11] version of MISITcms, that reduces the effort almost to a one-click solution. The different components are organized with docker-compose into different containers, including the underlying database to be initialized on startup.

The source code for the MISITcms is published with the open-source license GPLv3 and can be found on Github at https://github.com/frankkramer-lab/MISITcms-app.

# Results

We employed our platform in three practical courses over several semesters and thereby gradually improved MISITcms. The course was aimed at teaching basic and advanced Python programming skills and was focused on implementing biomedical applications.

## Experience from a Teacher Perspective

The application of automated evaluation of student submissions turned out to be a great improvement in contrast to previous courses, which depended on the manual evaluation. Since the exercise had not to be checked by a teacher for each student individually, not only the course could be held with less staff, but also the obtained time could be spent on providing better individual support for the students.

After a short initial period, in which the students required support in interpreting eventually occurring error messages, the feedback provided by our platform was sufficient to enable them to work largely independently.

Each course day had a dedicated topic for which the corresponding materials were provided on our platform. New topics were activated on the day firstly needed, which then facilitated the start into the new topic and lead them directly to address the exercises.

Additionally, the reduced workload affected a reduced stress level of the tutors during the course, which was generally appreciated. However, this was paid for by a one-time work overhead during the preparation phase, since the exercises not only needed to be prepared but also integrated into the platform. Choosing a suitable specificity of the test cases was particularly one of the persistent problems. A simple comparison of the results of the user submissions to a predefined solution turned out to be sufficient in many cases, but often it was necessary to create custom tests and even refine those over the courses. In other cases, ambiguous solutions or rounding errors had to be taken into account, which also leads to more sophisticated checking.

Especially in the first attempts of the course, we experienced technical problems with the SAE server,

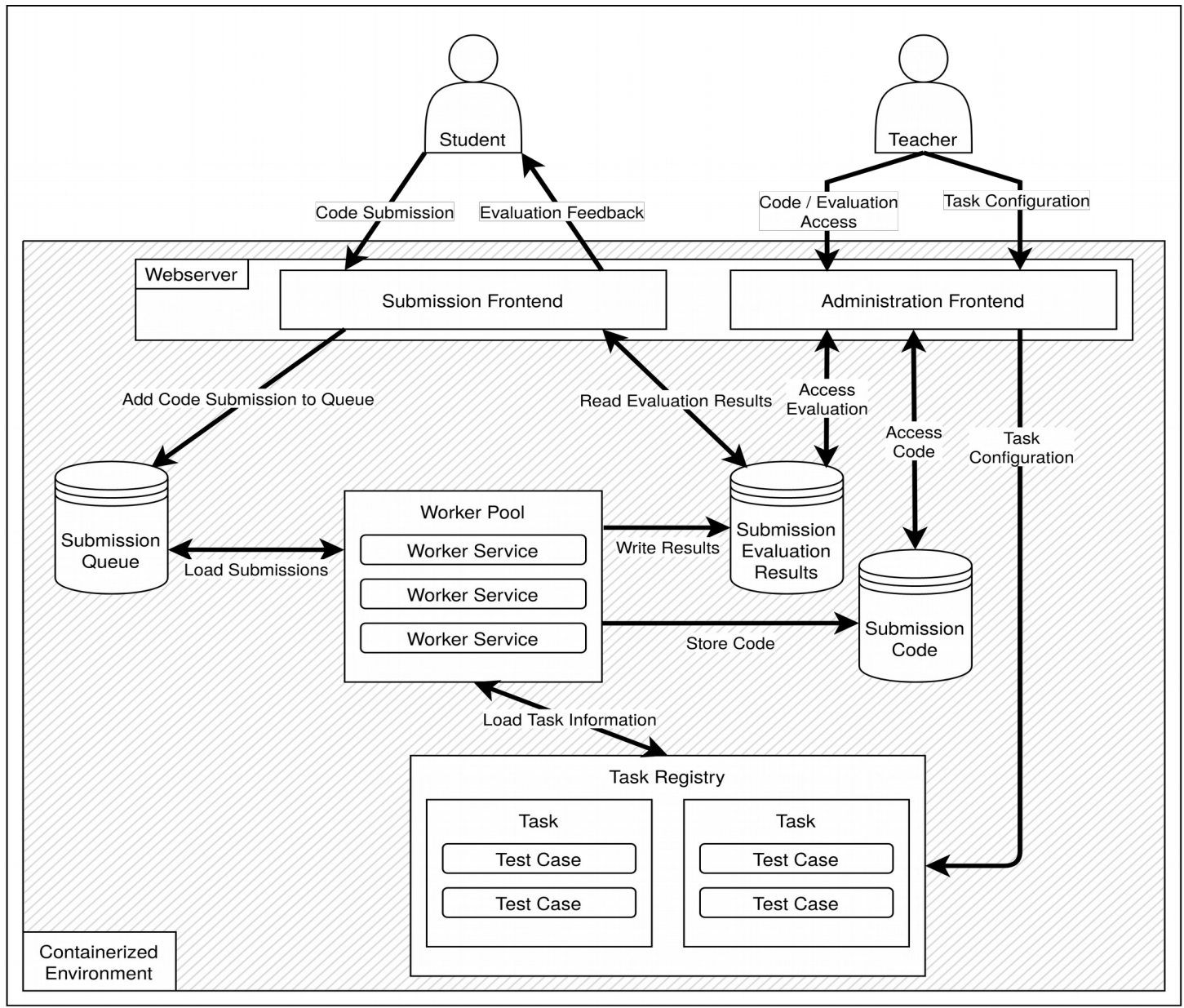

*Figure 2 – Containerized architecture of MISITcms illustrating the interplay of the different components.*

which initially lead to our advancements. A stable IT infrastructure is crucial for a smooth flow of the course, but still, on some occasions in the beginning this couldn't be provided. Therefore it is particularly important that our improvements guaranty more stability and advanced error handling.

## Students' Experience

Besides our own experiences, we were eager for feedback from the participants of the practical courses. Therefore, we designed a retrospective survey to capture their point of view on the features of our platform in terms of functionality, usability, transparency and feedback, the code submission process compared to other practical courses, and the retention of the SAE platform for this course. The survey was conducted anonymously and out of the 25 students, who completed the course, 9 (36%) participated in the survey. The results of the survey are illustrated in Figure 3.

While the basic functionality of our platform was perceived as commonly satisfying, the usability shows a clear trend towards being more positively accepted. The transparency of the evaluation of the tasks was experienced general impartial, with a slight tendency to the positive side, whereas the automated feedback on the results and error messages seems to drift towards the contrary.

The question of how the students assess the procedure of submitting the programming code, and its evaluation process, compared to other programming courses can not be answered clearly. Although the majority is quite neutral or even positive towards an SAE process, the opinions spread from great approval to a preference of traditional methods.

Nevertheless, the majority of the participants recommended the further usage of our platform for this practical course or at least supported the recommendation. Only one student expressed refusal towards the usage of an SAE platform within this course.

However, the low response rate to the survey, in addition to the readily comprehensible number of participants of the course lowers the expressiveness to some extent.

## Future improvements

Our own experience, as well as the feedback of the students, show, that a further enhancement of the test case specificity and more verbose feedback messages to the students are required. This will help to tackle the shortcomings reported for transparency and automated feedback.

Additional improvements in functionality and usability MISITcms could be extended towards a platform for code execution and interactive computation, including Jupiter Notebook style features. Since our platform already acts as a repository of the course materials, even further extension towards a massive open online course (MOOC) [12] is imaginable.

An important topic not addressed yet within MISITcms, and SAE platforms, in general, is checking for plagiarism. Therefore an automated validation could be included, which compares the submitted code against the submission of other students to determine shared fragments or whole blocks of code. This can be extended even further to a check against common online resources like Stack Overflow or GitHub to avoid blindly copy-and-pasting third-party solutions.

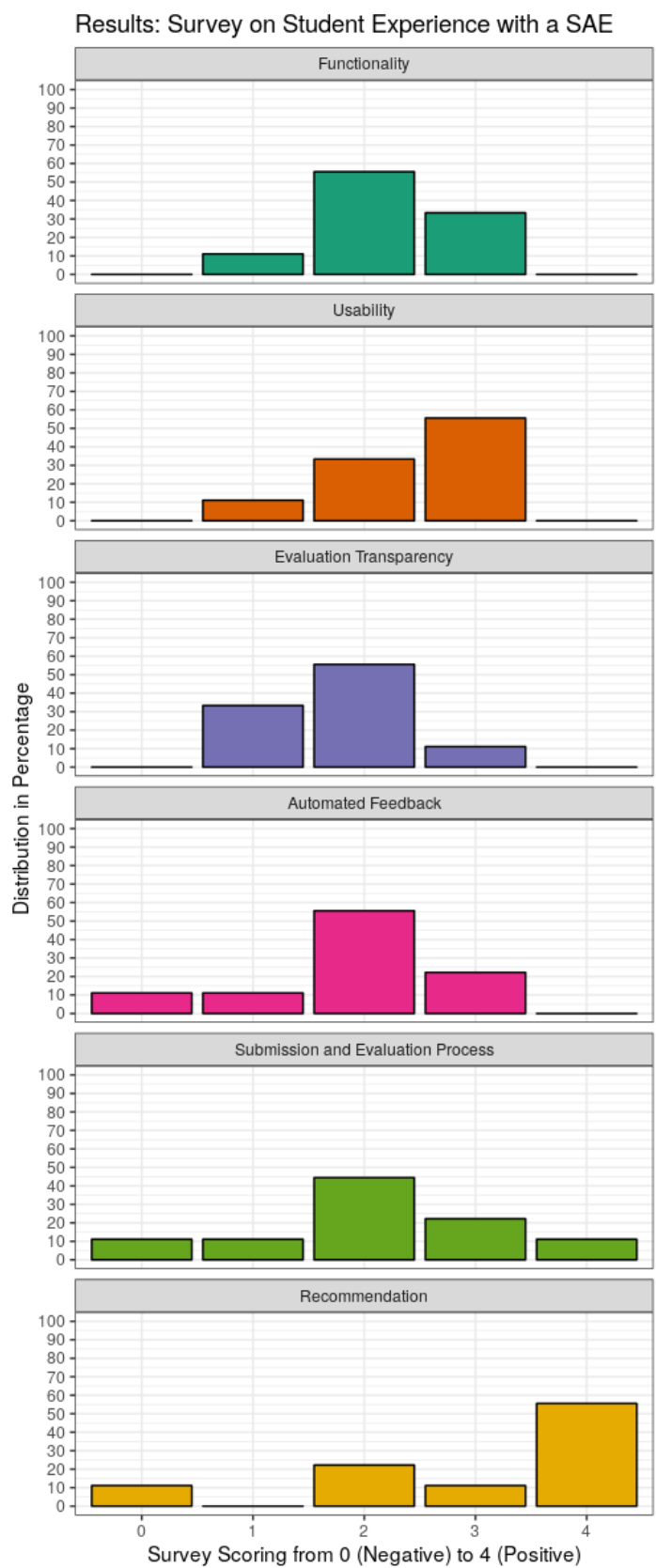

*Figure 3 – Evaluation results of the retrospective student survey on their perspective on SAE from a programming course in university.*

## Conclusions

In this work, we introduced an implementation of a code submission and automated evaluation server for student programming assignments and discussed positive and negative aspects of these systems for university teaching. Our implementation MISITcms integrated key features like a student-friendly web interface and learning environment, scalable and secure code execution environments, dynamic integration of teacher-defined tasks, support for multiple programming languages, and automated feedback for students.

Through utilizing MISITcms for university teaching as well as conducting a retrospective student survey over the last two years, we were able to identify multiple positive and negative aspects. The core benefits were the robust, transparent as well as automated code evaluation reducing the workload of teacher grading, the scalability, and the utilization of a structured as well as powerful platform for coding courses. MISITcms can act as a hub for course materials and also provides capabilities for archiving code submissions. However, we encountered challenges like an additional work overhead for creating automation-suitable assignments, handling or accepting various solution strategies and possible coding errors as well as the requirement of a stable IT infrastructure.

In summary, we concluded that an automated submission and evaluation platform for programming assignments requires continuous development and adjustment, but allows highly robust, structured as well as scalable student code evaluation which reduces the workload of teachers and which resulting transparency is appreciated by students.

## Acknowledgments

This work is a part of the DIFUTURE project funded by the German Ministry of Education and Research (Bundesministerium für Bildung und Forschung, BMBF) grant FKZ01ZZ1804E.

Manuscript accepted for publication in the following Conference - Jorunal: MedInfo 2021, Virtual Conference, October 2-4, 2021 - IOS Press.

## Address for correspondence

Dominik.Mueller@informatik.uni-augsburg.de